# Growth and site-specific organization of micron-scale biomolecular devices on living mammalian cells

Sisi Jia[1], Siew Cheng Phua[2], Yuta Nihongaki[2], Yizeng Li[3,§], Michael Pacella[1], Yi Li[1], Abdul M. Mohammed[1], Sean Sun[3], Takanari Inoue[2], Rebecca Schulman[1,4*]

[1] Chemical and Biomolecular Engineering, Johns Hopkins University, Baltimore, Maryland 21218, USA.

[2] Cell Biology, Johns Hopkins University School of Medicine, Baltimore, Maryland 21205, USA.

[3] Mechanical Engineering, Johns Hopkins University, Baltimore, Maryland 21218, USA.

[4] Computer Science, Johns Hopkins University, Baltimore, Maryland 21218, USA.

[§] Current address: Department of Mechanical Engineering, Kennesaw State University, Marietta, GA 30060

[*] To whom correspondence can be addressed. E-mail: rschulm3@jhu.edu

**Abstract:**
Mesoscale molecular assemblies on the cell surface, such as cilia and filopodia, integrate information, control transport and amplify signals. Synthetic devices mimicking these structures could sensitively monitor these cellular functions and direct new ones. The challenges in creating such devices, however are that they must be integrated with cells in a precise kinetically controlled process and a device's structure and its precisely structured cell interface must then be maintained during active cellular function. Here we report the ability to integrate synthetic micron-scale filaments, DNA nanotubes, into a cell's architecture by anchoring them by their ends to specific receptors on the surfaces of mammalian cells. These filaments can act as shear stress meters: how anchored nanotubes bend at the cell surface quantitatively indicates the magnitude of shear stresses between 0-2 dyn/cm$^2$, a regime important for cell signaling. Nanotubes can also grow while anchored to cells, thus acting as dynamic components of cells. This approach to cell surface engineering, in which synthetic biomolecular assemblies are organized within existing cellular architecture, could make it possible to build new types of sensors, machines and scaffolds that can interface with, control and measure properties of cells.

**Introduction:**

Micron-scale molecular assemblies, including membrane-bound[1] and membrane-less organelles[2], cilia[3], cytoskeletal networks[4] or the glycocalyx[5], spatially organize living cells, create specialized reaction environments, serve as transport conduits, and amplify chemical and mechanical signals in ways individual molecules cannot. Assembling synthetic micron-scale cell structures and controlling their dynamics are key goals of synthetic biology and nanotechnology[6] because these abilities could make it possible to construct, for example, new cellular reaction chambers, sensors, and information and material conduits.

A key challenge in this pursuit is that the formation and evolution of the cell's architecture is primarily kinetically driven[4]. The time-dependent concentrations of the assembling species and must be controlled to direct where and how many structures are assembled and how long they persist, and thus to build dynamic structures that integrate functionally into a cell's constantly evolving architecture.

A cell's architecture extends from its interior to its surface. Organizing and directing molecules on the cell surface is important for controlling cell fate[7], drug and gene delivery, and building biotic-abiotic interfaces[8] such as by attaching nanoparticles, small molecules[9] and nanowire cell-electronic interfaces[10] to the cell surface. While controlling interactions between cell receptors and nanostructures has been studied in the context of therapeutic modulation of receptor activity[11,12], or for directing import of therapeutics[13-16], less is known about creating and organizing microstructures that programmatically modify and extend cell surface architecture[7,17].

Micron-scale filaments are ubiquitous cell motifs that serve as sensors (antennae)[18], mechanical supports, agents for generating motion[19] or for transport. Filaments must grow and be anchored in prescribed orientations to execute these functions and actively grow and reorganize to maintain their structure and respond to stimuli. Here we organize micron-scale filaments that can act as functional cellular elements on specific cell surface receptors (Figure 1a). We then grow these anchored filaments, demonstrating their capacity for dynamic reorganization. We also demonstrate how the cell-anchored filaments are sensitive flow rate meters whose dynamic range encompasses physiologically relevant rates of blood or ion channel-activated[20,21] flow. We thus show how micron-scale structures can be attached to a cell at specific locations in specific orientations, extending the functional mesoscale architecture of the cell.

**Results:**

We used DNA tile nanotubes (Figure 1b), semiflexible filaments with persistence length 8.7±0.5 µm[22] (on order that of actin[23]) that polymerize via Watson-Crick hybridization[24,25]. These DNA nanotubes can grow from DNA origami templates, seeds (Figure 1c)[22], and can reach 100 µm in length[24-26]. DNA nanotube growth kinetics[25,27-30], hierarchical assembly pathways[31] and diffusion rates[22] have also been extensively characterized, allowing kinetic control over their growth and interactions with cells. Nanotubes can be functionalized with polymers, gold nanoparticles[32], proteins[33] and peptides[34], and thus could be templates for constructing diverse functional devices.

We sought an approach for anchoring DNA nanotube ends to specific receptors on living cells that could be easily tailored to target different receptor types. The design of such an anchoring process presents key challenges. First, a nanometer-scale anchor point on a filament's end must binding specifically to the chosen receptor[35], and the filament's much larger remaining surface must not interact with the cell. The nanotube's anchoring rate must also be higher than its rate of detachment or cell import[36]. Microparticles can be anchored to cells because microparticles' large surface areas allow high net attachment rates[37,38]; molecules or complexes can be reliably anchored when they are supplied at high concentrations (>>10 nM)[35]. Anchoring nanotubes requires interaction with a small area of nanotube surface, and because of DNA nanotubes' large size (~50 megadaltons), it is only practical to present them at concentrations <100-200 pM. To overcome these challenges, we developed a method in which a DNA nanotube seed serves as an anchor and presents numerous binding sites that attach quickly and effectively irreversibly at the desired receptor. This approach yields efficient attachment to multiple receptors on multiple cell types with little nonspecific binding.

**Reliably anchoring nanotube seeds to cell receptors.** We first characterized and eliminated nonspecific interactions between DNA nanotube seeds and nanotubes and cells[39]. We measured the rate of DNA nanotube seed/cell interaction by adding Atto488-labeled DNA nanotube seeds (final concentrations 8-64 pM) to HeLa cells in culture (Supp. Note S5). Confocal micrograph z-stacks showed that the average fluorescence intensity of seeds at the cells' midline increased linearly with seed concentration, with 107±17 attached seeds per cell for 64 pM seeds (Supp. Note S6, Supp. Figure S4a, b).

Poly(ethylene) glycol (PEG) coating can reduce nonspecific interactions between nanoparticles and cell membranes[40]. To test whether PEG coating might reduce nonspecific interaction between DNA nanostructures and cells, we hybridized 20 kD PEG-15 nt DNA strand conjugates to seeds (Figure 1c). Almost no PEG-coated seeds were visible on cells after 8-64 pM PEG-coated seeds were incubated with HeLa cells (Supp. Figure S4c, d).

We then conjugated 20 kD PEG to nanotube monomers (Figure 1d) and prepared seeded nanotubes by combining 415 nM PEG-conjugated monomers with 37 pM seeds and incubating them at 37°C in TAE-Mg$^{2+}$ buffer for 3 days. >40±4.8% of the resulting filaments were >3 µm long (Supp. Figure S5). Neither PEG-coated nanotubes grown from PEG-coated or unmodified nanotube seeds attach to cells (Supp. Note S8, Supp. Figure S6).

We first tried anchoring DNA nanotube seeds to cells using the SpyTag peptide and SpyCatcher protein, which form a covalent bond[41]. We hybridized six SpyTag peptide-DNA conjugates (Supp. Note S9 and Supp. Figure S7 a, b) to each seed's barrel[22]. We then expressed a GFP-integrin-SpyCatcher fusion protein in HeLa cells *via* transfection (Supp. Note S10). However, almost no nanotubes grew from SpyTag-modified seeds attached to cells (Supp. Note S11 and Supp. Figure S7c, d), perhaps because of the low SpyTag-SpyCatcher reaction rate constant[41]: 1400 ± 40 M$^{-1}$S$^{-}$

1. Even assuming fusion receptor overexpression ($10^4$ per cell)[42], at 64 pM nanotubes on average just one nanotube would anchor to each cell per hour (Supp. Note S12). Anchoring nanotubes requires a much faster binding reaction, so we next considered antibody-receptor interactions, as most protein interactions have forward rate constants of $10^5$-$10^6$ M$^{-1}$s$^{-1}$.

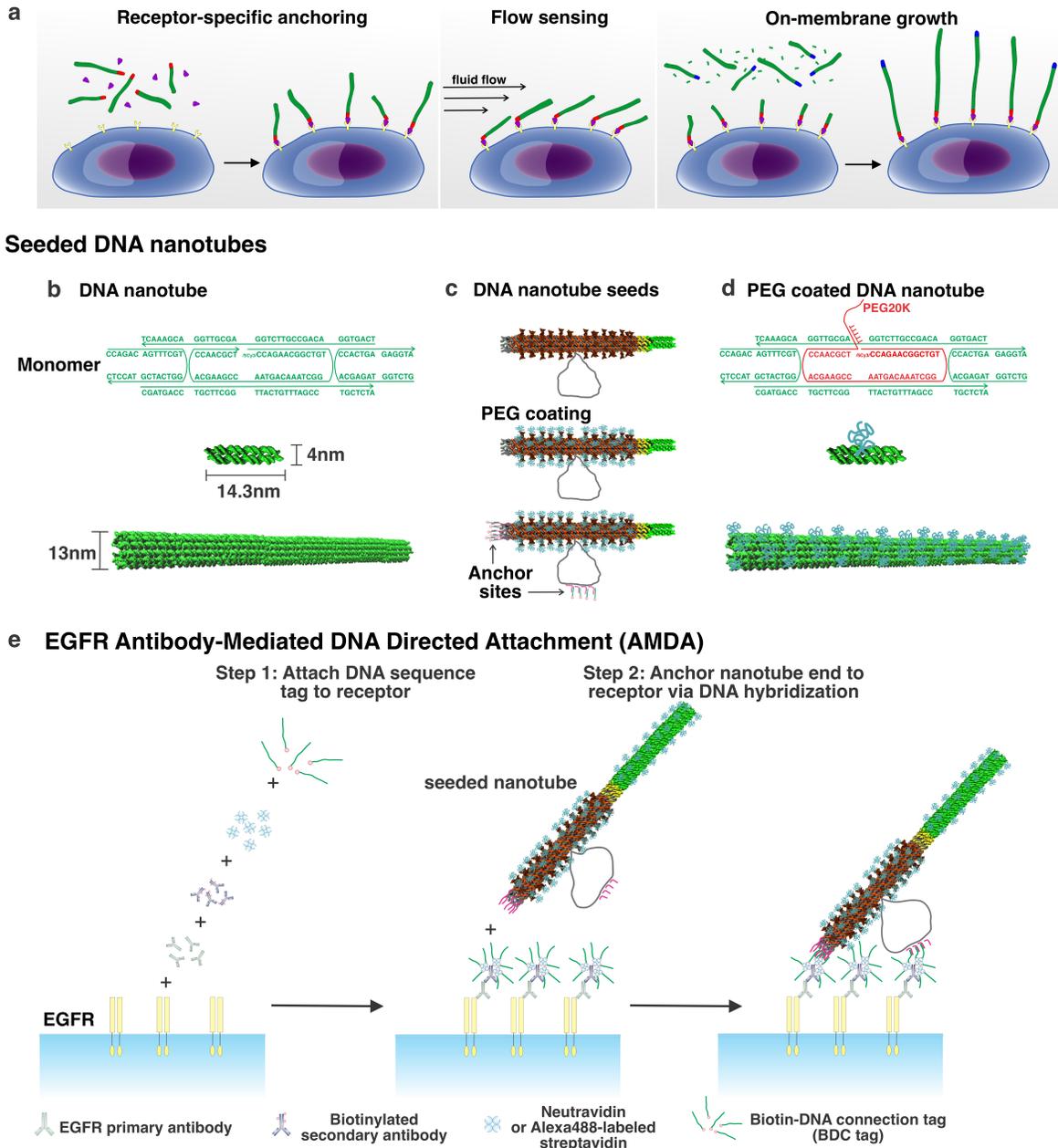

**Figure 1: Anchoring synthetic filaments, DNA nanotubes, to specific cell surface receptors. a.** DNA nanotubes anchored at specific locations on the cell surface could act as dynamic, functional elements of cells. **b.** Micron-scale DNA nanotubes self-assemble from monomer complexes. Arrows indicate 3' ends of DNA strands. **c.** Nanotube seeds are scaffolded DNA origami structures that template nanotube growth. Seeds can be coated with DNA-Polyethylene glycol (PEG, molecular weight 20 kD) conjugates (Supp. Figure S1 and S2). **d.** PEG-coated DNA nanotube monomers and assembled nanotubes. **e.** Schematic of EGFR antibody-mediated, DNA-Directed Attachment (AMDA) for anchoring seeded DNA nanotubes to cell surface receptors to origami seeds on nanotube ends. Primary antibodies, biotinylated secondary antibodies, streptavidin or neutravidin, and biotinylated DNA form complex to present a DNA sequence. This sequence hybridizes to the complementary DNA sequence on a DNA nanotube seed.

We tried to anchor DNA nanotube seeds to epidermal growth factor receptors (EGFR) using EGFR-EGFR antibody binding. EGFR is a transmembrane receptor tyrosine kinase overexpressed at up to $10^6$ copies per HeLa cell[43]. As expected, fluorescently labeled secondary antibodies attached to fixed and live HeLa cells only after EGFR primary antibodies were added (Supp. Note S13-S15, Supp. Figure S8). However, only a few DNA origami seeds with six secondary antibodies on their barrels[22] attached per cell (Supp. Note S16, Supp. Figure S9). We realized that while receptor-antibody binding was likely fast enough for hundreds of seeds to attach, most antibodies (including the EGFR antibody[44]) have >nM affinities[35]. Provided at picomolar concentrations, DNA origami seeds would thus not remain attached on average.

We thus turned to DNA hybridization. Forward rate constants[45] of DNA hybridization are $10^5$-$10^6$ $M^{-1}s^{-1}$, and a 15-nt DNA strand binds to its complement with sub-picomolar affinity under physiological conditions (Supp. Note S17). We designed a process in which a DNA sequence, termed the biotin-DNA connection (BDC) tag, is first attached to a receptor using antibodies, biotin and neutravidin. Nanotube seeds then present the BDC tag complement (BDC'), which hybridizes to the tethered BDC (Figure 1e). This approach can easily be generalized to attach different structures to different cell receptors: different antibodies could present different BDC sequences and different nanostructures their respective complements. We termed this scheme "antibody-mediated, DNA-directed attachment" (AMDA).

To test AMDA, we first added >10 nM each of primary EGFR antibodies, biotinylated secondary antibodies, neutravidin and finally BDC tag in steps to live HeLa cells (Supp. Table S10). We then added 16 or 64 pM nanotube seeds. About 2-fold more seeds attached to cells after either 16 pM or 64 pM seeds presenting 6 sequences complementary to BDC, BDC', at their barrels' ends than after a control AMDA process where the BDC strand was not added (Supp. Note S19, Supp. Figure S10).

We hypothesized that seeds' PEG coating might cover the BDC' sequences. We added 24 thymines to the BDC' presenting strands to increase the distance between the BDC' sequence and the PEG. We also replaced 30 fluorescently labeled DNA strands attached to a loop of DNA on the seed (Figure 1c) with strands presenting the BDC' sequence. These changes dramatically increased the number of seeds attached to HeLa cells after AMDA without increasing nonspecific attachment (Figure 2a, Supp. Note S20). Cross-sectional images showed seeds on the cell membrane, consistent with receptor attachment (Figure 2b). Elimination of any AMDA step almost completely eliminated seed attachment (Figure 2c, Supp. Figure S11).

We next used AMDA to attach nanotube seeds to EGFR on suspended HEK293 cells (Supp. Note S23). Nanotube seeds were present all over cells after AMDA, while little attachment was observed in controls (Figure 2d). The fluorescence intensity over background of HEK293 cells as determined by flow cytometry (674±75) was >5-fold higher after seeds were attached by AMDA than after a control process (125±55) (Figure 2f, Supp. Figure S12 and Supp. Note S24).

To verify that DNA seeds attached proximal to EGFR, we measured the colocalization of nanotube seeds with fluorescently labeled EGFR antibodies (Figure 2g, h). 76±4% (N=12 cells) of seeds were colocalized with EGFR antibodies after AMDA; Stochastic attachment would result in only 20±2% (N=12 cells) colocalization (Figure 2i, Supp. Note S25).

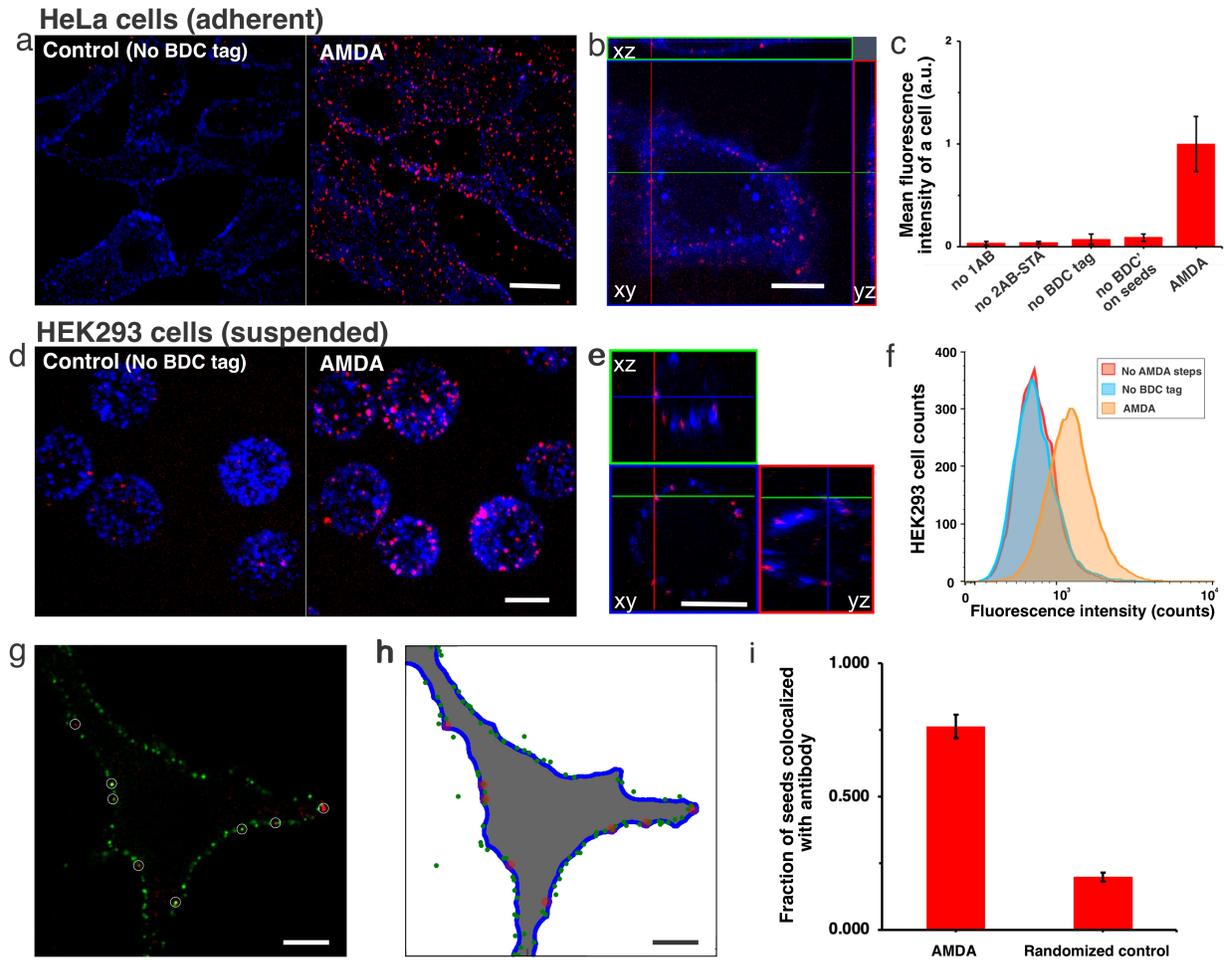

**Figure 2 PEG-coated DNA nanotube seeds attach to EGFR receptors via AMDA. a, d.** Three-dimensional projection images of HeLa (**a**) and HEK293 (**d**) cells after AMDA or AMDA with BDC tag addition omitted to attach seeds to EGFR (Supp. Note S20 and S23). Nanotube seeds were labeled with Atto488 (red) and secondary antibody-streptavidin conjugates with Alexa647 (blue). Scale bars: 20 μm. **b.** Confocal micrograph cross-sections of HeLa cells stained with DiD dye (blue) (Supp. Note S21) before seeds (red) were attached via AMDA. Scale bar: 20 μm. **c.** Average fluorescence intensities of nanotube seeds per HeLa cell after AMDA or after AMDA omitting different reagents (Supp. Note S22). **e** Confocal micrograph cross-sections of HEK293 cells. Scale bar: 20μm. **f** HEK293 cell fluorescence in the channel used to label seeds after AMDA (orange), AMDA with BDC tag addition omitted (blue) and no AMDA (red), measured via flow cytometry (Supp. Note S24). Average fluorescence intensities were 1423±75 (N=9818), 883±55 (N=9836) and 759±6 (N=9867). **g-i.** Co-localization of nanotube seeds and antibodies on the cell membrane (Supp. Note S25). **g.** Confocal micrograph cross-section of a HeLa cell after AMDA with secondary antibody labeled with Alexa647 (green) and nanotube seeds labeled with atto488 (red). Scale bar: 10μm. **h.** Computerized localization of antibodies (green) and seeds (red) of cell in g. **i.** Fractions of nanotube seeds colocalized with antibody after AMDA and in randomized controls. Error bars here and elsewhere, unless otherwise stated, are 95% confidence intervals.

Seeded DNA nanotubes attached reliably to HeLa cell membranes via AMDA but not in controls (Figure 3a-c, Supp. Note S26). Cells were 35-fold more fluorescent in the nanotube seed channel (Atto488) and 45-fold more fluorescent in the nanotube (Cy3) channel over background after AMDA vs. after a control (Supp. Note S27). Seeds did not move in time-lapse movies, but attached nanotubes moved freely (Supp. Movie S1), indicating that nanotubes were anchored to cells by seeds. Nanotubes could also be anchored to EGFR on HEK293 cells (Figure 3d-f, Supp. Movie S2 and Supp. Note S28). And nanotubes could be anchored to integrin receptors on HeLa cells *via* AMDA, demonstrating AMDA's generality (Supp. Note S30 and Supp. Figure S13).

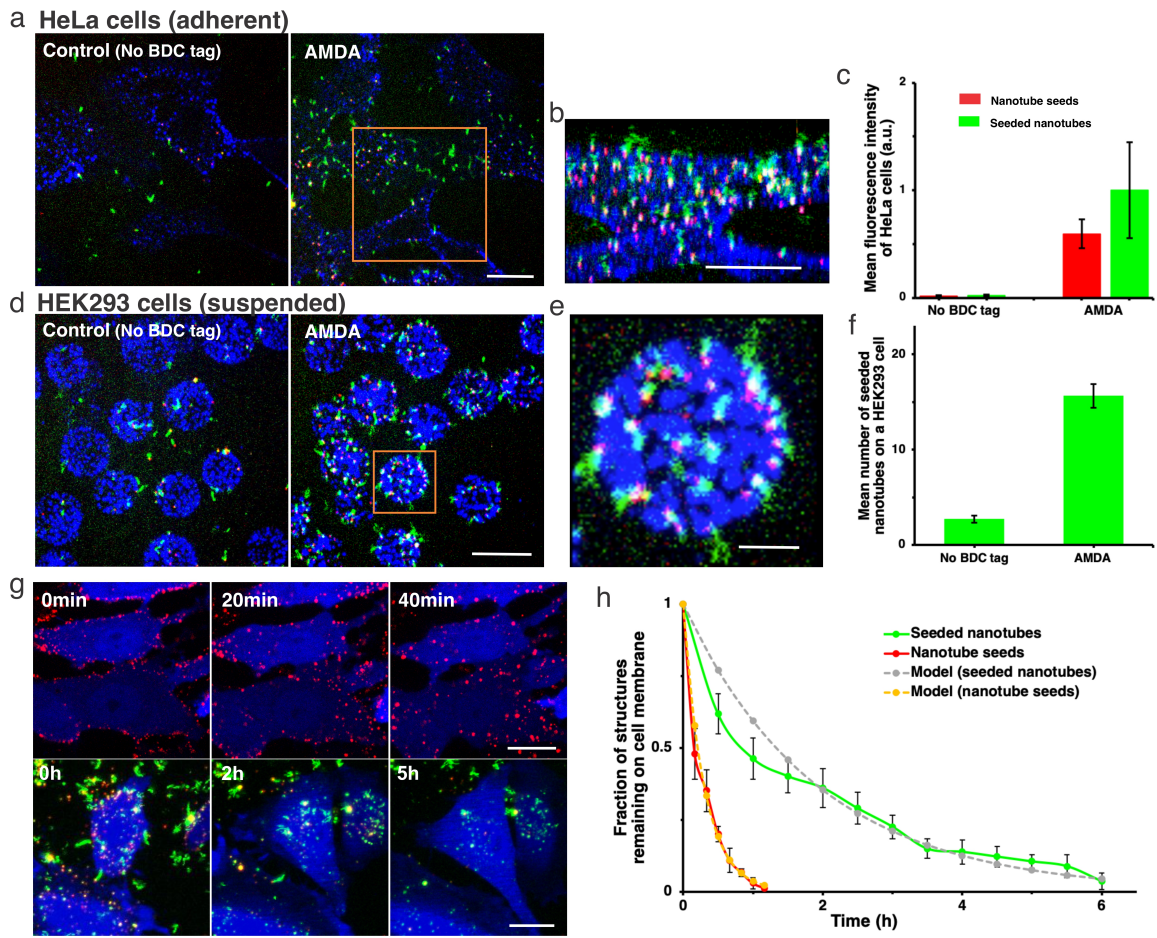

**Figure 3. Anchoring DNA nanotubes to EGFR receptors using AMDA. a** Confocal micrographs of seeded nanotubes anchored to HeLa cells after EGFR AMDA and EGFR AMDA with BDC tag addition omitted (Supp. Note S26). Seeds labeled with atto647 (red) and nanotubes with Cy3 (green), streptavidin with Alexa488 (blue). Scale bar: 20 μm. **b** Three-dimensional reconstruction of a HeLa cell with seeded nanotubes attached to EGFR. Scale bar: 20 μm. **c** The average fluorescence intensities of HeLa cells with seeded nanotube or seeds attached via AMDA and AMDA with BDC tag addition omitted (Supp. Note S27). **d** Three-dimensional projection images of HEK293 cells with attached seeded nanotubes after AMDA and AMDA with BDC tag addition omitted (Supp. Note S28). Scale bar: 20 μm. **e** Three-dimensional reconstruction of an HEK293 cell with attached seeded nanotubes. Scale bar: 5 μm. **f** Average numbers of seeded nanotubes on HEK293 cells after AMDA and after AMDA with BDC tag addition omitted (Supp. Note S29). Average numbers of attached seeds are shown in Figure 2f. **g.** Confocal micrographs of HeLa cells at different times after seed attachment using EGFR AMDA (upper panel) and maximum projection images of HeLa cells at different times after seeded nanotube attachment using EGFR AMDA (lower panel) (Supp. Notes S31, S32). Scale bar: 20 μm. **h** Fractions of DNA origami seeds or seeded nanotubes remaining on the cell surface at different times after AMDA (N=6 cells for each case) (Supp. Note S33). Fits are (a*  For seeds, a=1.0±0.002, b=3.3±0.50 /h, for seeded nanotubes a=1.0±0.002, b=0.52±0.06 /h.

We next asked how long nanotubes or nanotube seeds would persist on a cell's surface at 37 ºC, as receptor turnover or seed endocytosis could lead to detachment or import. DNA origami seeds and seeded nanotubes were first attached to HeLa-GFP cells at 4 °C, where detachment rates were low. The cells were then returned to 37 °C where decreases in the number of seeds or nanotubes on the cell surface were measured using time-lapse confocal microscopy (Supp. Notes S31 and S32). The fractions of nanotubes and nanotube seeds on the surface both decreased exponentially with time (Figure 3g-h). The time constant for seeds was 6 times faster than for seeded nanotubes (Figure 3h).

These measurements did not distinguish whether structures detached from or were imported into the cell[46]. EGFR-mediated endocytosis is a receptor-mediated clathrin-dependent pathway[47] in which a membrane invagination is pinched off by the motor protein dynein. Nanotube seeds (Figure 1d, length:65nm) are small enough to conceivably be endocytosed with EGFR[48]. EGFR-mediated endocytosis takes on order 30 minutes[49], consistent with the rate of seeds leaving the cell surface. Nanotubes are too large to be endocytosed, but dynein-controlled membrane closure might sever them. EGFR is a fast-turnover receptor[50] and HeLa cells are fast-growing cells so these persistence times are likely at the lower range across different receptors and cell lines.

**Nanotube shear stress sensors.** We next tested whether anchored nanotubes could measure cell surface shear stress. Shear can result from flow and is a key environmental signal *in vivo*. For example, the primary cilium[3] is involved in sensing flow in the kidney[51], and bends in response to flows[21,52,53], inducing signaling[54,55]. We asked whether, like primary cilia, nanotubes might bend in response to shear stress and whether the extent of this bending might indicate the magnitude of shear stress.

To assess this possibility, we developed a model of how a nanotube anchored to the surface of a rectangular chamber would respond to shear stress induced by laminar flow of velocity $U$ (Figure 4a, Supp. Note S34). A nanotube was modeled as a rigid rod anchored by a flexible linker. The chamber was much taller (400 μm) than a nanotube's length, so the flow field around the nanotube should be essentially uniform (Figure 4b). In simulations, the polar angle between the nanotube and z-axis was close to $\pi/2$ except at very small shear stresses, so we assumed this polar angle was $\pi/2$ under external flow (see Supp. Note S34.2, Supp. Figure S19). The nanotube's response could therefore be reduced to an in-plane (xy plane) rotation, *i.e.* the azimuth angle, $\phi$ between the nanotube and the flow's direction (Figure 4b). In this case, the flow-induced viscous drag on the nanotube is $\mathbf{F} = (\alpha\mu U\ell, 0)$, where $\alpha$ is the coefficient of viscous drag on the nanotube, $\mu$ is the viscosity of the fluid in the chamber, and $\ell$ is the nanotube's length. The directional vector of the center of mass of the nanotube is $\mathbf{r} = (\ell/2 \cos\phi, \ell/2 \sin\phi)$. Thus, the torque on the nanotube is $\mathbf{M} = \mathbf{r} \times \mathbf{F} = -1/2\ \alpha\mu U\ell^2 \sin\phi$. We used this torque to calculate the dynamics of the nanotube, which are governed by $\gamma\ d\phi/dt = M + R$, where $\gamma$ is the nanotube's damping coefficient, $M = |\mathbf{M}|$, and $R$ is a random force from thermal fluctuations. $R$'s distribution is given by $P(R) \propto \exp[-R^2\ \Delta t/(2k_BT\gamma)]$, where $\Delta t$ is the time step used to numerically evolve the equation. For each time step, a random $R$ was drawn. The initial value of the azimuth angle, $\phi_0$, of each nanotube was randomly drawn from the uniform distribution $[-\pi, \pi]$. We solved the probability distribution of $\phi$ by sampling $\phi$ for a large number of nanotubes for each a set of volumetric flow rates $Q = UHW$, where $H$ and $W$ are, respectively, the chamber's height and width (Figure 4c). We found that the distributions of azimuthal angles should vary for shear stresses between 0-1.5 dyn/cm$^2$, a range relevant for ion channel activation[20,21].

To measure the sensitivity and dynamic range of nanotube flow sensors, we anchored nanotube seeds to the bottom of a passivated glass microchannel[22] (Supp. Note S36 and S37) and measured their orientations under different flows using time-lapse spinning disk confocal microscopy (Figure 4d, Supp. Notes S38). In the absence of flow, nanotubes explored all azimuthal angles and bent in the z-direction. A shear stress of only 0.05 dyn/cm$^2$ caused the nanotubes to remain in plane and align with the flow (Figure 4d). To quantify the relationship between nanotube orientation and fluid shear stress on glass, we measured the mean total angle of nanotube rotation over 30 frames taken every 5 seconds at different fluid shear stresses (Supp. Note S41). A maximum time projection image of each nanotube was generated from these images indicating the nanotube's total angular range of angular motion, $\Phi$ (Figure 4f). The mean $\Phi$ for different nanotubes experiencing a given shear stress, $\overline{\Phi}$, decreased with increasing shear stresses between 0.05-2 dyn/cm$^2$, consistent with our model's predictions (Figure 4h).

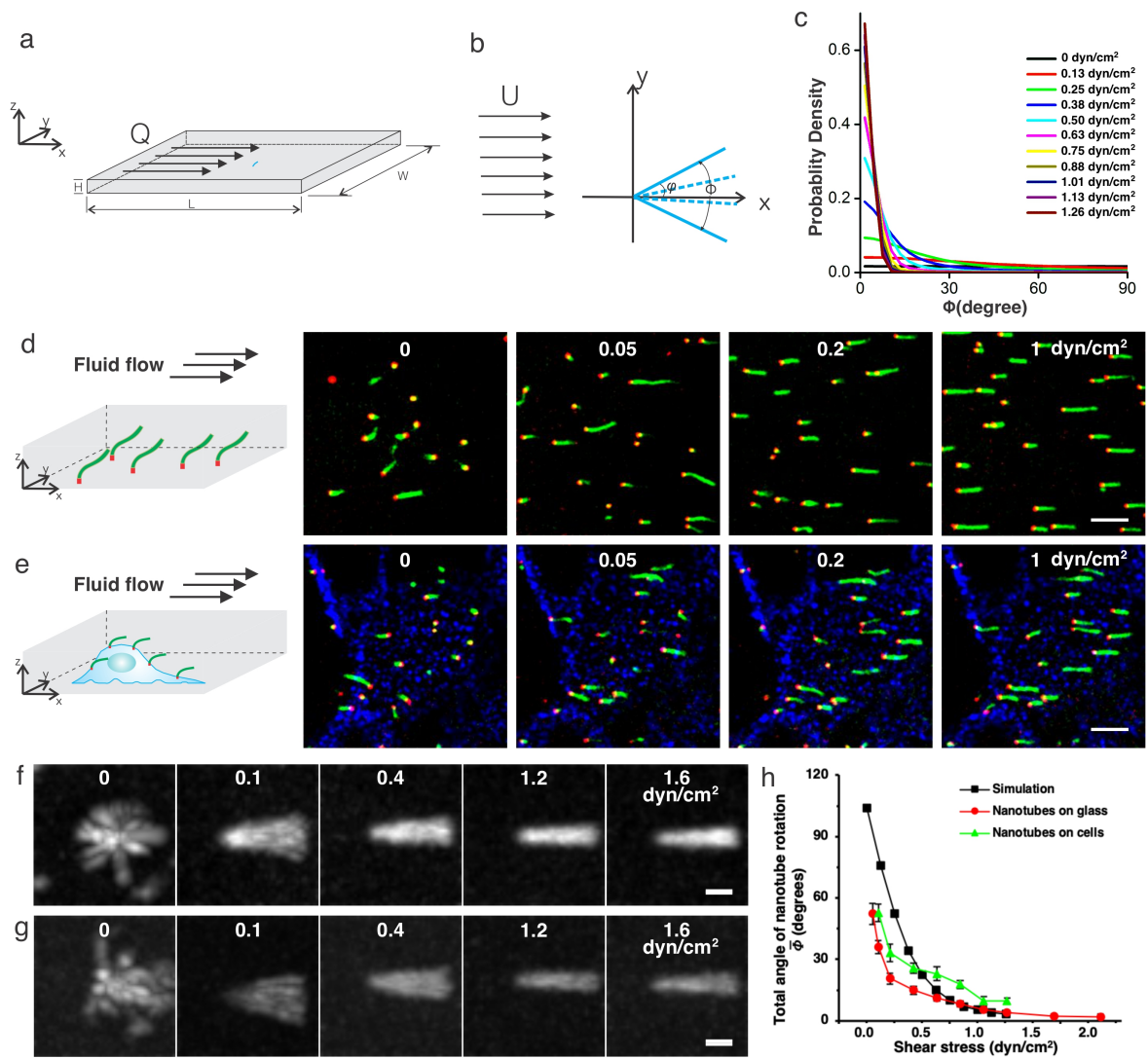

**Figure 4 Anchored nanotubes indicate the magnitude of shear stress at the cell surface**. **a,b** Side (a) and top (b) views of nanotube deflection in a flow Q in a rectangular channel. φ is the azimuthal angle between the plane of the nanotube and the x-axis. $\Phi$ is the total angle of nanotube rotation over a given time duration. **c** Predicted distribution of φ by a simple model of deflection (Supp. Note S44.3). **d, e** Confocal micrographs of seeded nanotubes anchored on the glass surface of a rectangular flow chamber (height=0.54mm, width=3.8mm) (**d**) and the top of HeLa cell membranes (**e**) in response to fluid shear stresses of 0, 0.05, 0.2, and 1 dyn/cm$^2$. Nanotubes were labeled with Cy3 (green), nanotube seeds with atto647 (red), and the cell membrane visualized with streptavidin-Alex488-conjugated EGFR antibodies (blue). Scale bars: 10 μm. **f, g** Maximum projection images of seeded nanotubes anchored on glass (**f**) and on HeLa cells (**g**) in response to fluid shear stresses 0, 0.1, 0.4, 1.2 and 1.6 dyn/cm$^2$. Scale bars: 2 μm. **h** Mean total angles of nanotubes as a function of fluid shear stress. N= 15 nanotubes for each shear stress on both glass and cells.

Nanotubes attached to cells via AMDA (Supp. Notes S39 and S40) also increasingly aligned with the flow as shear stress increased (Figure 4e, g and Supp. Movie S3). Because cells are not flat, a nanotube's location on a cell affected its bend direction and motion. The total angles of rotation of nanotubes on the tops of cells varied most in response to different flow rates. $\overline{\Phi}$ for nanotubes on the tops of HeLa cells was close to both the values predicted by the model and the values measured on glass for all shear stresses, suggesting how anchored nanotubes can serve as "windsocks" on cells that indicate flow direction and the magnitude of shear stress the flow induces.

**Growing nanotubes on living cells**. A key advantage of using self-assembled biomolecular structures as cell surface microdevices is that they might dynamically grow or reorganize *via* biomolecular reactions. To explore the possibility of constructing such dynamic microdevices, we asked whether DNA nanotubes could grow while anchored to cell receptors.

Nanotubes can grow via monomer addition but at monomer concentrations where end-on growth is preferred over homogeneous nucleation, growth occurs at <0.2 μm/h[22,29,56]. Because nanotubes persist only a few hours on EGFR, we sought instead to extend nanotubes via end-to-end joining of pre-assembled nanotubes[26,57]. While rapid end-to-end DNA nanotube joining has been observed *in vitro*[22], only 7±2% (N=477) of PEG-coated nanotubes underwent end-to-end joining within 4 hours in cell buffer at physiological temperatures (Supp. Note S42 and Supp. Figure S23). We hypothesized that end-to-end joining did not occur because the monomer detachment rate was very low, allowing rough facets or facets with defective monomers that cannot join to persist (Supp. Figure S25). To increase the monomer detachment rate, we shortened the monomers' binding sites from 6 to 4 nucleotides[58] to produce 4PEG nanotubes. 86±3% (N=803) of 4PEG nanotubes anchored to glass surface grew *via* end-to-end joining within 3.5 hours after 4PEG nanotubes (green) and 150 nM monomers that could serve as "glue" to fill in gaps between rough facets (red) were added (Supp. Note S45 and Supp. Figure S26).

60±9% of 4PEG nanotubes attached to EGFR on HeLa cells were extended using a similar protocol of end-to-end joining and monomer gluing (Figure 5a-c, Supp. Movie S4 and Supp. Note S46). Since the fluid flow used to determine whether individual nanotubes had grown sometimes severed them (Supp. Figure S27), we repeated the joining process and then added methylcellulose to reduce nanotube diffusion (Supp. Note 48). This process revealed alternating green-red segments indicating nanotube gluing and joining (Figure 5c-e, Supp. Figure S29 and Supp. Movie S5) as well as overlapping red and green segments indicating filament bundling, which high viscosity medium can induce[59].

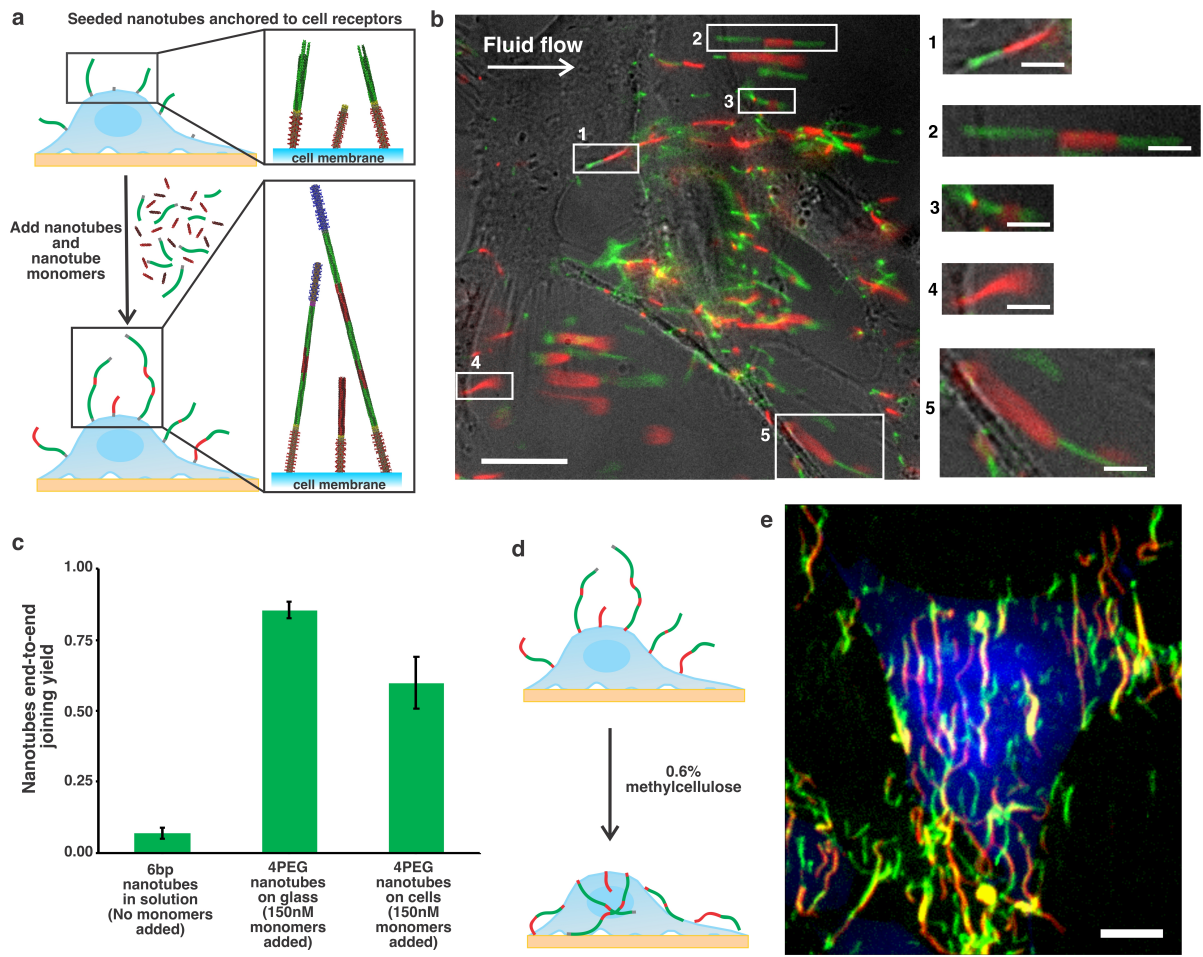

**Figure 5: Nanotube growth on the surfaces of living cells. a** Schematic of nanotube growth via end-to-end joining and monomer addition. Seeds on anchored and capped nanotubes were unlabeled. Nanotubes were labeled with Cy3 (green), monomers with atto647 (red). **b** Two-color fluorescence micrograph of joined nanotubes anchored to live HeLa cells. Gentle fluid flow (shear stress 0.32 dyn/cm$^2$) was applied to stretch the nanotubes for better characterization (Supp. Note S46). Scale bar: 20 μm. Zoom-in images of end-to-end joined nanotube structures on the cell surface. Scale bar: 5 μm. **c** Nanotube end-to-end joining yield quantification: 6nt seeded nanotubes incubated with capped seeded nanotubes in solution at 37°C for 3.5h without additional monomers (Supp. Note S42 and Supp. Figure S23), 4PEG seeded nanotubes anchored on glass surface incubated with capped nanotubes and additional 150nM monomers added at 20°C for 3.5h (Supp. Note S45 and Supp. Figure S26), 4PEG seeded nanotubes anchored on cell membrane incubated with capped nanotubes and additional 150nM monomers at 20°C for 4h (Supp. Note S46). **d-e** Schematic (**d**) and 3D projection images (**e**) of joined nanotubes (green and red) on a live HeLa cell after 0.6% methylcellulose (IMDM) was added (Supp. Note S48). Cells were transfected with GFP (blue) to reveal cell shape and extent. Scale bar: 10 μm.

## Conclusions

While biomolecular filaments are structurally simple, they can assemble in myriad ways to create complex functional materials and devices, as exemplified by cytoskeletal structures. Here we site-specifically anchored synthetic DNA filaments to living cells by determining the required binding affinities, reaction rates and avidity for efficient attachment, and mitigating nonspecific interactions. The resulting precise control over attachment, in combination with our understanding of DNA nanotube growth rates[25,29], hierarchical assembly[31] and reorganization[57], might be used to build a range of synthetic, dynamic filament-based devices on cells, including antennae, motion-inducing devices or conduits that connect receptors on different cells. More generally, the understanding of how filament binding kinetics and thermodynamics, assembly timing, and component stoichiometry affect filament growth and organization might also enable the design of genetically encoded processes to direct the assembly of filaments synthesized by the cells themselves.

## Methods

**Reagents**. M13mp18 scaffold strand was purchased from Bayou Biolabs. All other DNA strands used in this study were synthesized by Integrated DNA Technologies, Inc. (IDT). Strands for the DNA nanotube tiles (Supp. Note S49) and the adapter strands for the DNA nanotube seeds (Supp. Note S50.3), Cy3-, ATTO647- and ATTO488-, biotin-labeled strands and amino-modified strands were HPLC purified. All other strands were simply desalted. Concentrations of DNA strands were determined either by measuring absorbance at 260 nm (using extinction coefficients supplied by IDT) or by relying on IDT to determine solution concentrations. N-hydroxylsuccinimide (NHS) functionalized polyethylene glycol (molecular weight 20K) (PEG-20K) was purchased from NANOCS (PG1-SVA-20K). Phosphate buffered saline (PBS) (28372) was purchased from ThermoFisher and prepared at 10x for further use. Gel loading dye blue (B7021S) was purchased from New England Biolabs and Sybr gold (S11494) was purchased from ThermoFisher. Centrifugal filters (UFC510096) for purifying seeds were purchased from MilliporeSigma. For cell culture, HeLa cell and HEK293 cell lines were both purchased form ATCC. DMEM medium (10-013-CV) was purchased from Corning Cellgro. FBS (26140079), 1% penicillin-streptomycin (15140122), 0.05% Trypsin-EDTA (25300054) and DPBS (14190144) were all purchased from ThermoFisher. For cell transfection, the Opti-MEM(1x) (31985062) was purchased from ThermoFisher and the X-tremeGENE (6365779001) form Sigma Aldrich. The azide-modified SpyTag peptide (Lot No. P3130-1) was synthesized by BioSynthesis. The size exclusion spin column Illustra Microspin G-25 was purchased from GE Healthcare. For AMDA, EGFR monoclonal antibody (H11) (MA513070), the Alexa fluor 647(A21236)-conjugated secondary antibody, the biotin-conjugated secondary antibody (31800), the streptavidin-Alexa Fluor 488 conjugate (S32354), neutravidin (31000) and DiD live cell labeling solution (V-22887) were all purchased from ThermoFisher. Integrin β1 Antibody (K-20) (sc-18887) (for integrin AMDA) was purchased from Santa Cruz Biotechnology. Bovine serum albumin (BSA) (A3858) and $MgSO_4$ were purchased from Sigma-Aldrich. The Streptavidin Conjugation Kit (ab102921), used for conjugating streptavidin with Alexa 647 labeled secondary antibody, was purchased from Abcam. Borosilicate glass Lab-Tek 8-well chambers (155411PK) were purchased from ThermoFisher. Glass-bottom dishes (μ-Dish 35 mm, high Grid-50 glass bottom) (81148), μ-slide VI 0.4 (80606) and glass bottom μ-slide channel VI 0.5 (80607) were purchased from Ibidi. Biotin-PEG-silane (Biotin-PEG-SIL-3400−500 mg) was purchased from Layson Bio. Methylcellulose (HSC001) was purchased from R&D systems and Iscove's Modified Dulbecco's medium (IMDM) (12440053), which was used to dilute the methylcellulose, was purchased from ThermoFisher.

**Synthesis of PEG-DNA strand conjugates.** 8 mg NHS-PEG20k was dissolved in 100 μL of a PBS buffer solution (pH 7.2) containing 50 μM amino-modified DNA strand. The mixture was agitated at room temperature (19-20 °C) overnight to allow the reaction to run to completion. Afterwards, the solution containing the PEG20K-DNA conjugates was loaded into a 7% PAGE gel. The running buffer was TAE-$Mg^{2+}$ (40 mM Tris-acetate, 1 mM EDTA to which 12.5 mM magnesium acetate was added) and loading buffer 1x blue gel loading dye. The gel run at 150V for 1h. The desired band containing the PEG-DNA conjugate was cut out and the conjugate was extracted from the gel by soaking the gel in water for 2-4 days to let the conjugate diffuse out from the gel. The concentration of the PEG20K-DNA conjugate was determined by quantifying the amount of Cy3-labeled PEG-DNA conjugate using fluorescence intensity and using a known concentration to quantitate the unlabeled conjugate via PAGE gel (Supp. Note S2)

**Assembly of PEG-coated DNA nanotube seeds.** The structure and sequences of DNA nanotube seeds used in this work are described in Supp. Note S50. To create PEG-coated DNA nanotube seeds, the staple sequences of a DNA origami seed structure[25] were modified to each present a DNA sequence that served as an attachment site for a PEG-DNA conjugate (sequence AAGCGTAGTCGGATCTC). The resulting seeds were assembled, purified and their concentrations measured using protocols adopted from Agrawal *et al*[29] (Supp. Note S1 step 2 and 3). To coat the resulting nanotube seeds with PEG, 18μL of a solution containing 10 μM PEG-DNA conjugate and 1.8μL 10x TAE-$Mg^{2+}$ buffer were added to 100 μL of a TAE-$Mg^{2+}$ solution containing 0.8 nM seeds and incubated on the bench for 30min (see Supp. Note S3).

**Growing PEG-coated seeded nanotubes.** The structures and sequences of DNA nanotube monomers with both 6- and 4-base sticky end binding sites are given in Supp. Note S49. To grow PEG-coated seeded nanotubes with 6 base sticky ends, the central SEs3 strand of the tile was conjugated with PEG as described in Supp. Note S7. 19.7μL of a TAE-$Mg^{2+}$ solution containing 450 nM of each of the strands for the monomers were annealed from 90 to 37 °C as described in Supp. Note S1 step 2. 2 μL of a solution containing 0.4 nM PEG-coated seeds was added after the monomer solution reached 37 °C. The mixture was kept at 37 °C for 3 days. To grow 4PEG nanotubes, the central REd3 and SEd3 strands of the two monomer types were each conjugated with PEG. 19.7 μL of a TAE-$Mg^{2+}$ solution containing 180 nM of each the strands of the two 4PEG nanotube monomers was annealed from 90 to 20 °C as described in Agrawal *et al*[29]. 2 μL of a TAE-$Mg^{2+}$ solution containing either 0.4 nM PEG-coated anchored nanotube seeds (Supp. Table S20 and S21) or 0.4 nM PEG-coated capped nanotube seeds (Supp. Note S22) as appropriate for experiments on nanotube joining were added after the solution reached 20 °C. The solution was then incubated at 20 °C for 3 days.

**Cell culture.** HeLa cells and HEK 293 cells were grown in DMEM medium containing 10% FBS and 1% penicillin-streptomycin. 5mL of culture was grown in 25 $cm^2$ culture flasks at 37 °C in 5% $CO_2$ and constant humidity. Cells were released from the flask surface using 0.05% Trypsin-EDTA and split every two days. The HeLa cells were cultured in media with: 3, 6, 9, and 12 mM $MgSO_4$ overnight, after which time cell viability was confirmed by shape under a bright-field microscope (Supp. Note S4).

**Characterizing the extent of nonspecific interactions between DNA nanotube seeds or seeded nanotubes and HeLa cells.** HeLa cells were seeded in borosilicate glass Lab-Tek 8-well chambers at a density of 40000 cells per well in 250 μL medium (Supp. Note S5 step 2). DNA nanotube seeds with and without PEG coating were diluted to make 8, 16, 32 and 64 pM solutions in a cold DMEM solution containing 1% BSA (w/v) and 12mM $MgSO_4$. The medium in each well chamber containing HeLa cell was exchanged for 250 μL of diluted seeds solution. The cells were then incubated in a 4° C refrigerator for 30 minutes and subsequently washed with DMEM-12mM

MgSO$_4$ buffer 3 times (Supp. Note S5). Seeded DNA nanotubes grown from 37 pM seeds with PEG or 28 pM seeds without PEG in TAE-Mg$^{2+}$ were diluted one-fold with cold 1% BSA(DMEM)-12mM MgSO$_4$. The medium in the wells containing HeLa cells was exchanged for diluted nanotube solution. The cells were then incubated in a 4 ºC refrigerator for 2 hours and subsequently washed with DMEM-12mM MgSO$_4$ 3 times (Supp. Note S8). After both treatments, the cells were fixed using 4% paraformaldehyde before imaging.

**Transfection of HeLa cells with SpyCatcher-fusion transgenes.** The GFP-integrin-SpyCatcher plasmids were constructed by inserting the SpyCatcher DNA sequence into the GFP-integrin construct at the NotI restriction site. The backbone of the plasmid is a Clontech vector with kanamycin resistance. For plasmid sequences, see Supp. Note S51. The plasmid was transformed and amplified in DH5alpha bacteria, and amplified using a Qiagen miniprep kit. HeLa cells for transfection were cultured and passaged as described above. HeLa cells were counted using a hemacytometer and diluted to 2.4x10$^5$ cell per mL in DMEM. For each transfection process, 38 µL of Opti-MEM (1x) and 2µL of a solution containing 1 mg/mL plasmid DNA were mixed well in a 1.5mL Eppendorf tube *via* pipetting. 2 µL X-tremeGENE 9 solution was then added and the solution again mixed well *via* pipette, after which the mixture was incubated at room temperature for 30 minutes. 42 µL of the plasmid solution was then added to 500 µL of diluted cells in a 1.5 mL Eppendorf tube and mixed well by gently inverting the tube around 20 times. 250 µL of cell solution was then pipetted into a well of borosilicate glass Lab-Tek 8-well chamber. The cells were then incubated for 2 days before use (see Supp. Note S10).

**Attachment of seeded nanotubes to cells using SpyCatcher-SpyTag attachment.** SpyTag-modified seeded DNA nanotubes were prepared by attaching an azide-modified SpyTag peptide to an amino-modified DNA oligonucleotide *via* a click reaction (Supp. Note S9) and hybridizing this strand to the complementary sequence presented at the ends of nanotube seeds which were then spin-filtered to remove extra SpyTag (Supp. Note S1 step 3). Nanotubes were then grown from seeds with and without the SpyTag modification and diluted 1.5-fold with DPBS-12 mM MgSO$_4$ buffer. The medium in each well chamber containing HeLa cells expressed with GFP-integrin-SpyCatcher fusion protein was exchanged for 250 µL of diluted nanotube solution prepared as described for characterization of nonspecific nanotube-cell interactions but diluted by DPBS with 12 mM MgSO$_4$ buffer and incubated at 37 ºC/5% CO$_2$ for 30 minutes (Supp. Note S11).

**Attachment of nanotube seeds or seeded nanotubes to EGFR receptors on HeLa cells using AMDA.** PEG-coated nanotube seeds or seeded nanotubes were prepared as described as above and HeLa cell were seeded overnight in borosilicate glass Lab-Tek 8-well chambers at a density of 40000 cells per well with 250µL medium (Supp. Note S5 step 2). The next morning, the cells were first incubated in a 4 ºC refrigerator for 10 minutes. DMEM buffer containing 1% BSA (w/v) was then added to the cells, which were then incubated for 5 minutes. To attach nanotube seeds, 250 µL of solution containing 1) 2 µg/mL EGFR primary antibody, 2) 10 µg/mL Alexa 647-labeled secondary antibody-streptavidin conjugate, 3) 1 µM BDC tag and 4) BDC' tag-labeled nanotube seeds at concentration to achieve the stated concentration in cell solution were each added to the cells in the order listed. After each addition the cells were incubated in the refrigerator for 30 minutes then washed 3 times with cold DMEM (DMEM-12mM MgSO$_4$ after seeds or nanotubes were added) to remove reagent not attached to the cells. To attach seeded nanotubes, the Alexa 647-labeled secondary antibody solution was replaced by 250 µL solutions containing

2a) 500-fold diluted biotinylated secondary antibody followed by 2b) 3 µg/mL Alexa488-labeled streptavidin. After each addition cells were washed 3 times with cold DMEM buffer. To attach nanotubes to cells, the seed solution was replaced by a one-fold diluted solution of seeded nanotubes. This solution was incubated with cells for 2 hours, pipetting gently every 30 minutes. After incubation cells were washed with DMEM-12 mM MgSO$_4$ buffer 3 times.

**Attachment of nanotube seeds or nanotubes to EGFR receptors on HEK293 cells using AMDA.** PEG-coated nanotube seeds and seeded nanotubes were prepared as described as above and HEK293 cells were trypsinized and suspended to a concentration of $10^6$-$10^8$ cells per mL. The cells were centrifuged at 300RCF for 5 minutes, the supernatant was removed and the cells were resuspended in cold 1% BSA in DMEM. To attach nanotube seeds to suspended HEK293 cells, the cells were centrifuged and resuspended in a solution containing 1) 2 µg/mL EGFR primary antibody, 2) 10 µg/mL Alexa 647 2AB-STA 3) 1 µM BDC tag and 4) PEG-coated nanotube seeds in sequence. After each addition, the cells were incubated in a 4 ºC refrigerator for 30 minutes and pipetted-mixed every 15 minutes, then washed by centrifuging and resuspending in 1 mL cold DMEM buffer to remove unattached reagent. After this sequence, the cells were resuspended in DMEM-12 mM MgSO$_4$ buffer. To attach seeded nanotubes, resuspension in Alexa 647-labeled secondary antibody solution was replaced by resuspension in 2a) 500-fold diluted biotinylated secondary antibody then 2b) 3 µg/mL Alexa488-labeled streptavidin. After the addition of the BDC tag and resuspension, the cells were resuspended in 1 mL cold DMEM buffer then a solution of seeded nanotube solution (1-fold diluted after preparation). The cells were incubated in a 4 ºC refrigerator for 2 hours during which time they were pipetted gently every 30 minutes. The cells were centrifuged a last time, then resuspended in DMEM-12 mM MgSO$_4$ buffer.

**Attachment of nanotubes to integrin receptors on HeLa cells using AMDA.** The steps for AMDA were followed above except that the EGFR primary antibody solution was replaced with a solution containing 4 µg/mL Integrin β1 antibody. After addition, cells were incubated at 4 °C for 1 hour.

**Spinning disk confocal microscopy.** Cells with attached DNA nanotube seeds or seeded nanotubes were imaged using a Zeiss AxioObserver Yokogawa CSU-X1 spinning disk confocal microscope with a 60x oil objective. Stack images were taken from the bottom of the cell to the top of the cell with a stack depth of 0.27µm (for HeLa cells) or 0.5 µm (for HEK293 cells) at 10-15 random locations.

**Quantification of the number of nanotube seeds and seeded nanotubes attached to each cell.** The average fluorescence intensity per cell was used to quantify the average number of seeds/seeded nanotube attached to a HeLa cell. A z-stack of images was collected at each imaging position, and the stack image from the height closest to the center of an average-sized cell was selected for analysis. This choice was made because this image largely excluded the structures attached to the glass rather than the cell while maintaining a sufficiently large cross-sectional area for each cell for analysis. The number/total length of nanotube seeds and seeded nanotubes present was measured by characterizing the total fluorescence intensity (Supp. Note S22 and S27). The seeds' intensity per cell and nanotube intensity per cell were then calculated by dividing these respective quantities by the number of cells in an image, which was also counted manually[60]. Flow cytometry (BD FACSCanto) was used to characterize the number of nanotube seeds on the

HEK293 cells (Supp. Note S24). The number of seeded nanotubes on HEK293 cells was determined by generating a 3-dimensional projection image from each of a z-stack of images collected at multiple random locations and manually counting the number of seeded nanotubes visible on each cell (Supp. Note S29). The amount reported is the average of these counts.

**Quantification of the dwell time of seeds and seeded nanotubes on the cell membrane after AMDA.** PEG-coated DNA nanotube seeds or seeded nanotubes were anchored on HeLa-GFP cells using EGFR AMDA as described above. The cells were washed with cold (4 ºC) DMEM-12mM $MgSO_4$ then placed into an incubator (37°C, 5% $CO_2$ and constant humidity) on a Nikon A1 confocal microscope (Nikon, Tokyo, Japan) with a 63× oil objective. Stacks of images at heights spanning the bottoms and tops of the cell in the field of view. Image stacks of cells with attached nanotube seeds were collected every 10 minutes over 70 minutes with a stack height of 0.27 μm. Image stacks of cells with attached nanotubes were collected every 15 minutes over 20 hours with a stack height of 1 μm. The number of nanotube seeds and seeded nanotubes on each tracked cell's membrane were counted manually. These numbers were normalized by the number of seeds or nanotubes on that cell at t=0. Seeds were counted if they were visible at a cell's edge. The top and bottom images of a stack were omitted from quantification because it was not possible to determine whether seeds were beneath or above the cell rather than at the cell surface. The total number of seeded nanotubes on a cell's surface at each time point was manually counted using maximum projection images (Supp. Note S33).

**Use of fluid flow to apply shear stress at a glass surface or HeLa cell membrane.** Shear stress was applied within in flow cells (Ibidi, μ-slide VI 0.5 for glass-anchored nanotubes and μ-slide VI 0.4 for cell-anchored nanotubes) and the shear stress corresponding to a given flow rate was calculated according to the methods provided by Ibidi (Supp. Note S35 and Supp. Table S17). Seeded nanotubes were anchored to the glass bottoms of flow cells using a method developed previously[22] (Supp. Note S36 and S37). A syringe pump (New Era, NE-1000) was used to induce controlled, unidirectional laminar flow. TAE-$Mg^{2+}$ buffer was used as flow perfusate (Supp. Note S38). Seeded nanotubes were anchored to the HeLa cell membrane through EGFR AMDA performed in a flow cell (Supp. Note S39). DMEM-12mM $MgSO_4$ buffer was used as flow perfusate (Supp. Note S40). Seeded nanotubes under fluid flow were imaged using a spinning disk confocal microscope with 5 seconds intervals for 30 cycles. The nanotubes were imaged in the xy-plane in which the largest number of seeds and the largest fraction of the nanotubes were in focus. The total angle of the nanotube rotation under fluid flow was measured by first cropping the area spanned by a single nanotube from each larger image. A Gaussian blur filter (radius:1.00) was applied in ImageJ for all the 30 cropped images to reduce the image background. A maximum time projection image was then generated from this cropped time-lapse movie and the total angle of the nanotube rotation under fluid flow was measured manually (Supp. Note S41). The total angle 15 nanotubes on the glass surface and >15 nanotubes on cell membrane was measured for each shear stress.

**Growth of seeded nanotubes anchored to the cell membrane.** 4PEG nanotubes and 4PEG capped and seeded nanotubes were prepared as described in Supp. Note S43. Here both the capped and anchored nanotube seeds were unlabeled (Supp. Table S20 and S22). Anchored seeded nanotubes were attached to the HeLa cell membrane through EGFR AMDA within a μ-slide channel. 50 uL of a solution containing 900 nM inactive nanotube monomers (Supp. Note S44) labeled with atto647 (Supp. Table S25) was annealed in TAE-Mg$^{2+}$ buffer from 90 to 20 °C. 0.27 µL of solution containing 100 µM of a strand to activate the inactive monomers (Supp. Note 46) was added to this solution after which 25 µL of it was immediately mixed with 60 µL of a solution containing 37 pM of capped nanotubes and 65 µL of DMEM-12.5mM MgSO$_4$ buffer containing 1% BSA. This mixture was immediately added to the HeLa cells to which seeded nanotubes had been attached. The sample was covered with foil and incubated on the lab bench (at about 19-21°C) for 4h. It was then washed with DMEM-12mM MgSO$_4$ three times before imaging. A gentle fluid flow (0.18 mL/min) inducing a shear stress of 0.32 dyn/cm$^2$ was applied to stretch the nanotubes and allow visualization of their contours. An epi-fluorescence microscope with a 60x oil objective was used to capture continuous 20 images of each location. The yield of nanotube joining on cell membrane was calculated by counting the total number of seeded nanotubes on the cell and the number of nanotubes on the cell that had visually joining. Cells in six images were quantified. 0.6% methylcellulose media (IMDM) with 12 mM MgSO$_4$ was added to HeLa GFP cells after the joining protocol (but not flow or imaging) was completed (Supp. Note S48). Stacks of images at random locations were taken from the bottoms to tops of cells with a stack height of 0.27 µm using a spinning disk confocal microscope.

**Acknowledgements:**
The authors thank Samuel Schaffter, Yi Li, and Pepijn Moerman for helpful discussion, Kostas Konstantopoulos, Panagiotis Mistriotis and Kaustav Bera for technical assistance and HeLa-GFP cell lines, Michael McCaffrey and Erin Prycell for imaging assistance. S.J., Yi.L. and R.S. acknowledge support from DARPA BTO Award D16AP00147 (YFA) and NSF CMMI-1562661. This work was supported in part by the US National Institutes of Health (NIH) grant to T.I. (DK102910). S.C.P. was supported by the Agency for Science, Technology and Research (Singapore). Y. N. was supported by postdoctoral fellowships from Japan Society for the Promotion of Science and from the Uehara Memorial Foundation. Yizeng L. and S.S. acknowledge support from NIH R01GM134542.

**Author Contributions:**
This study was conceived by S.J., R.S. and T.I.. S.J. designed the experiments, carried out the experiments, and analyzed and interpreted the data. S.C.P. and A.M.M designed the SpyTag-SpyCatcher fusion binding approach for nanotube-cell receptor attachment. S.C.P. constructed the GFP-integrin-SpyCatcher and GFP-integrin plasmids DNA and helped with cell culture and cell transfection. Y.N. helped with the flow experiment setup. M.P. and Y.L helped with data analysis. Y.L. and S.S. designed the flow model, performed the simulation and analyzed the simulation results. S.J. and R.S. wrote the manuscript with input and edits from all authors.

**Competing interests Statement:**
The authors declare no competing interests.

**References:**

1    Lippincott-Schwartz, J., Roberts, T. H. & Hirschberg, K. Secretory protein trafficking and organelle dynamics in living cells. *Annu. Rev. Cell Dev. Biol.* **16**, 557–589 (2000).
2    Shin, Y. & Brangwynne, C. P. Liquid phase condensation in cell physiology and disease. *Science* **357**, eaaf4382 (2017).

3   Phua, S. C., Lin, Y.-C. & Inoue, T. An intelligent nano-antenna: Primary cilium harnesses TRP channels to decode polymodal stimuli. *Cell Calcium* **58**, 415-422 (2015).
4   Fletcher, D. A. & Mullins, R. D. Cell mechanics and the cytoskeleton. *Nature* **463**, 485-492 (2010).
5   Reitsma, S., Slaaf, D. W., Vink, H., van Zandvoort, M. A. M. J. & oude Egbrink, M. G. A. The endothelial glycocalyx: composition, functions, and visualization. *Pflug. Arch. Eur. J. Phy.* **454**, 345-359 (2007).
6   Elbert, D. L. Bottom-up tissue engineering. *Curr. Opin. Biotech.* **22**, 674-680 (2011).
7   Bettinger, C. J., Langer, R. & Borenstein, J. T. Engineering Substrate Topography at the Micro- and Nanoscale to Control Cell Function. *Angew. Chem. Int. Ed.* **48**, 5406-5415 (2009).
8   Feiner, R. & Dvir, T. Tissue-electronics interfaces: from implantable devices to engineered tissues. *Nat. Rev. Mater.* **3**, 17076 (2017).
9   Nel, A. E. *et al.* Understanding biophysicochemical interactions at the nano-bio interface. *Nat. Mater.* **8**, 543-557 (2009).
10  Fattahi, P., Yang, G., Kim, G. & Abidian, M. R. A Review of Organic and Inorganic Biomaterials for Neural Interfaces. *Adv. Mater.* **26**, 1846-1885 (2014).
11  Jiang, W., Kim, B. Y. S., Rutka, J. T. & Chan, W. C. W. Nanoparticle-mediated cellular response is size-dependent. *Nature Nanotechnol.* **3**, 145-150 (2008).
12  Fan, J., Wang, H.-H., Xie, S., Wang, M. & Nie, Z. Engineering Cell-Surface Receptors with DNA Nanotechnology for Cell Manipulation. *ChemBioChem* **21**, 282-293 (2020).
13  Yu, B., Tai, H. C., Xue, W., Lee, L. J. & Lee, R. J. Receptor-targeted nanocarriers for therapeutic delivery to cancer. *Mol. Membr. Biol.* **27**, 286-298 (2010).
14  Dreaden, E. C., Austin, L. A., Mackey, M. A. & El-Sayed, M. A. Size matters: gold nanoparticles in targeted cancer drug delivery. *Ther. Deliv.* **3**, 457-478 (2012).
15  Habibi, N., Kamaly, N., Memic, A. & Shafiee, H. Self-assembled peptide-based nanostructures: Smart nanomaterials toward targeted drug delivery. *Nano Today* **11**, 41-60 (2016).
16  Charoenphol, P. & Bermudez, H. Aptamer-Targeted DNA Nanostructures for Therapeutic Delivery. *Mol. Pharm.* **11**, 1721-1725 (2014).
17  Teramura, Y. & Iwata, H. Cell surface modification with polymers for biomedical studies. *Soft Matter* **6**, 1081-1091 (2010).
18  Vogel, V. & Sheetz, M. Local force and geometry sensing regulate cell functions. *Nat. Rev. Mol. Cell Biol.* **7**, 265-275 (2006).
19  Pollard, T. D. & Cooper, J. A. Actin, a Central Player in Cell Shape and Movement. *Science* **326**, 1208-1212 (2009).
20  Davies, P. F. Flow-mediated endothelial mechanotransduction. *Physiol. Rev.* **75**, 519-560 (1995).
21  Su, S. *et al.* Genetically encoded calcium indicator illuminates calcium dynamics in primary cilia. *Nat. Methods.* **10**, 1105-1107 (2013).
22  Mohammed, A. M., Šulc, P., Zenk, J. & Schulman, R. Self-assembling DNA nanotubes to connect molecular landmarks. *Nat. Nanotechnol.* **12**, 312-316 (2016).
23  Janmey, P. A. *et al.* The mechanical properties of actin gels. Elastic modulus and filament motions. *J. Biol. Chem.* **269**, 32503-32513 (1994).

24   Rothemund, P. W. K. *et al.* Design and Characterization of Programmable DNA Nanotubes. *J. Am. Chem. Soc.* **126**, 16344-16352 (2004).
25   Mohammed, A. M. & Schulman, R. Directing Self-Assembly of DNA Nanotubes Using Programmable Seeds. *Nano Lett.* **13**, 4006-4013 (2013).
26   Axel Ekani-Nkodo, A. K., Deborah Kuchnir Fygenson. Joining and Scission in the Self-Assembly of Nanotubes from DNA Tiles. *Phys. Rev. Lett.* **93**, 268301 (2004).
27   Schulman, R. & Winfree, E. Synthesis of crystals with a programmable kinetic barrier to nucleation. *Proc. Natl. Acad. Sci.* **104**, 15236-15241 (2007).
28   Barish, R. D., Schulman, R., Rothemund, P. W. K. & Winfree, E. An information-bearing seed for nucleating algorithmic self-assembly. *Proc. Natl. Acad. Sci.* **106**, 6054-6059 (2009).
29   Agrawal, D. K. *et al.* Terminating DNA Tile Assembly with Nanostructured Caps. *ACS Nano* **11**, 9770-9779 (2017).
30   Schaffter, S. W., Green, L. N., Schneider, J., Subramanian, H K K., Schulman, R., Franco, E. T7 RNA polymerase non-specifically transcribes and induces disassembly of DNA nanostructures. *Nucleic Acids Res.* **46**, 5332–5343 (2018).
31   Jorgenson, T. D., Mohammed, A. M., Agrawal, D. K. & Schulman, R. Self-Assembly of Hierarchical DNA Nanotube Architectures with Well-Defined Geometries. *ACS Nano* **11**, 1927-1936 (2017).
32   Sharma, J. *et al.* Control of Self-Assembly of DNA Tubules Through Integration of Gold Nanoparticles. *Science* **323**, 112-116 (2009).
33   Hariadi, R. F. *et al.* Mechanical coordination in motor ensembles revealed using engineered artificial myosin filaments. *Nat. Nanotechnol.* **10**, 696 (2015).
34   Stephanopoulos, N. *et al.* Bioactive DNA-Peptide Nanotubes Enhance the Differentiation of Neural Stem Cells Into Neurons. *Nano Lett.* **15**, 603-609 (2015).
35   Landry, J. P., Ke, Y., Yu, G.-L. & Zhu, X. D. Measuring affinity constants of 1450 monoclonal antibodies to peptide targets with a microarray-based label-free assay platform. *J. Immunol. Methods* **417**, 86-96 (2015).
36   MacGurn, J. A., Hsu, P.-C. & Emr, S. D. Ubiquitin and Membrane Protein Turnover: From Cradle to Grave. *Annu. Rev. Biochem.* **81**, 231-259 (2012).
37   Singh, R. & Lillard, J. W. Nanoparticle-based targeted drug delivery. *Exp. and Mol. Pathol.* **86**, 215-223 (2009).
38   Cedervall, T. *et al.* Understanding the nanoparticle–protein corona using methods to quantify exchange rates and affinities of proteins for nanoparticles. *Proc. Natl. Acad. Sci.* **104**, 2050-2055 (2007).
39   Koyfman, A. Y., Braun, G. B. & Reich, N. O. Cell-Targeted Self-Assembled DNA Nanostructures. *J. Am. Chem. Soc.* **131**, 14237-14239 (2009).
40   Verma, A. & Stellacci, F. Effect of Surface Properties on Nanoparticle-Cell Interactions. *Small* **6**, 12-21 (2010).
41   Zakeri, B. *et al.* Peptide tag forming a rapid covalent bond to a protein, through engineering a bacterial adhesin. *Proc. Natl. Acad. Sci.* **109**, E690-E697 (2012).
42   Benedetto, S. *et al.* Quantification of the expression level of integrin receptor αvβ3 in cell lines and MR imaging with antibody-coated iron oxide particles. *Magn. Reson. Med.* **56**, 711-716 (2006).

43	Bremer, E. *et al.* Simultaneous Inhibition of Epidermal Growth Factor Receptor (EGFR) Signaling and Enhanced Activation of Tumor Necrosis Factor-related Apoptosis-inducing Ligand (TRAIL) Receptor-mediated Apoptosis Induction by an scFv:sTRAIL Fusion Protein with Specificity for Human EGFR. *J. Biol. Chem.* **280**, 10025-10033 (2005).

44	Zhou, Y. *et al.* Impact of Intrinsic Affinity on Functional Binding and Biological Activity of EGFR Antibodies. *Mol. Cancer Ther.* **11**, 1467-1476 (2012).

45	Schreiber, G., Haran, G. & Zhou, H. X. Fundamental aspects of protein-protein association kinetics. *Chem. Rev.* **109**, 839-860 (2009).

46	Wang, P. *et al.* Visualization of the Cellular Uptake and Trafficking of DNA Origami Nanostructures in Cancer Cells. *J. Am. Chem. Soc.* **140**, 2478-2484 (2018).

47	Vieira, A. V., Lamaze, C. & Schmid, S. L. Control of EGF Receptor Signaling by Clathrin-Mediated Endocytosis. *Science* **274**, 2086-2089 (1996).

48	Zhang, S., Li, J., Lykotrafitis, G., Bao, G. & Suresh, S. Size-Dependent Endocytosis of Nanoparticles. *Adv. Mater.* **21**, 419-424 (2009).

49	Mickler, F. M. *et al.* Tuning Nanoparticle Uptake: Live-Cell Imaging Reveals Two Distinct Endocytosis Mechanisms Mediated by Natural and Artificial EGFR Targeting Ligand. *Nano Lett.* **12**, 3417-3423 (2012).

50	Sorkin, A. & Duex, J. E. Quantitative analysis of endocytosis and turnover of epidermal growth factor (EGF) and EGF receptor. *Curr. Protoc. Cell Biol.* **Chapter 15**, Unit-15.14 (2010).

51	Rosivall, L., Mirzahosseini, S., Toma, I., Sipos, A. & Peti-Peterdi, J. Fluid flow in the juxtaglomerular interstitium visualized in vivo. *Am. J. Physiol-Renal.* **291**, F1241-F1247 (2006).

52	Rydholm, S. *et al.* Mechanical properties of primary cilia regulate the response to fluid flow. *Am. J. Physiol-Renal.* **298**, F1096-F1102 (2010).

53	Young, Y. N., Downs, M. & Jacobs, C. R. Dynamics of the Primary Cilium in Shear Flow. *Biophys. J.* **103**, 629-639 (2012).

54	Nauli, S. M. *et al.* Polycystins 1 and 2 mediate mechanosensation in the primary cilium of kidney cells. *Nat. Genet.* **33**, 129-137 (2003).

55	Li, Q. *et al.* Polycystin-2 cation channel function is under the control of microtubular structures in primary cilia of renal epithelial cells. *J. Biol. Chem.* **281**, 37566-37575 (2006).

56	Hariadi, R. F., Yurke, B. & Winfree, E. Thermodynamics and kinetics of DNA nanotube polymerization from single-filament measurements. *Chem. Sci.* **6**, 2252-2267 (2015).

57	Green, L. N. *et al.* Autonomous dynamic control of DNA nanostructure self-assembly. *Nat. Chem.* **11**, 510-520 (2019).

58	Pacella, M. S. *et al.* Characterizing the length-dependence of DNA nanotube end-to-end joining rates. *Mol. Syst. Des. Eng.* (2020).

59	Popp, D., Yamamoto, A., Iwasa, M. & Maéda, Y. Direct visualization of actin nematic network formation and dynamics. *Biochem. Bioph. Res. Co* **351**, 348-353 (2006).

60	Wilkinson, M. H. F. & Schut, F. *Digital Image Analysis of Microbes: Imaging, Morphometry, Fluorometry and Motility Techniques and Applications*. (Wiley, 1998).